OPEN

# Quantum Search on Encrypted Data Based on Quantum Homomorphic Encryption

Qing Zhou[1,2,3], Songfeng Lu[2,3*], Yongquan Cui[2*], Li Li[4] & Jie Sun[1]

**We propose a homomorphic search protocol based on quantum homomorphic encryption, in which a client Alice with limited quantum ability can give her encrypted data to a powerful but untrusted quantum server and let the server search for her without decryption. By outsourcing the interactive key-update process to a trusted key center, Alice only needs to prepare and encrypt her original data and to decrypt the ciphered search result in linear time. Besides, we also present a compact and perfectly secure quantum homomorphic evaluation protocol for Clifford circuits, where the decryption key can be calculated by Alice with polynomial overhead with respect to the key length.**

Due to the great challenge of building large-scale quantum computers, it is very likely that only a few powerful quantum computers are initially available and act as quantum servers. Suppose Alice has some confidential data and wishes to search on them with the aid of a remote quantum server, Bob, but is unwilling to reveal the data to him. A natural approach to achieving this is to encrypt the data before handing them over and to let Bob directly search on the cipher-state, namely, to employ a quantum search scheme on encrypted data.

The classical counterpart of this issue has been studied extensively, where mainly two approaches are adopted: (1) devising a special cryptosystem supporting search operations and (2) utilizing a pre-built searchable cipher index offered by the client, both of which provide computational security rather than information-theoretic security[1]. In the quantum context, Sun proposed a quantum symmetric searchable encryption scheme for classical inputs[2], in which encryption and decryption are rotations along a coordinate axis, and the secret key composes of random rotation angles. The search procedure of this scheme is to linearly check and compare the encrypted items, which is inefficient for a large search space. Intuitively, linear search on encrypted data could be accelerated by Grover's algorithm[3], provided that there exists an encryption method allowing quantum operations without the encryption key. Fortunately, quantum homomorphic encryption (QHE) satisfies this requirement, which intrigued us to take advantage of its merits and to construct a quantum homomorphic search protocol.

The notion of QHE was first introduced by Liang[4] in 2013, who observed the commutation rules between the encryption operators of quantum one-time pad (QOTP)[5] and a universal set of quantum gates (containing rotation operators and CNOT gate) and demonstrated a symmetric one-party QHE scheme where homomorphic evaluations require the secret key. Later, he presented an interactive two-party scheme[6] based on commutation rules for another universal gate set (consisting of $H$, $S$, CNOT and $T$ gates). Aside from QHE, there are many other closely related constructions based on the combination of QOTP and commutation rules, such as secure assisted quantum computation[7], quantum computing on encrypted data[8], and delegating private quantum computations[9], which could be considered as early forms of QHE.

In 2014, Yu *et al.*[10] came up with a no-go result on QHE, which concludes that "quantum mechanics does not allow for efficient information-theoretically-secure fully homomorphic encryption scheme". In view of this result, Broadbent and Jeffery devised two remarkable $T$ gadgets for non-Clifford circuits and presented three efficient QHE schemes based on classical q-IND-CPA secure homomorphic encryption (CHE), one of which is designed for Clifford circuits and the other two are for non-Clifford circuits with a limited number of $T$ gates[11]. They also formally defined an important property—compactness—for QHE. In outline, a QHE scheme is compact if the circuit complexity of the decryption procedure for any allowed evaluation $C$ is independent of the circuit complexity of $C$; the decryption procedure is the task being performed by the client to recover the outcome evaluated by the server. For example, the interactive schemes[4,6–9] based on QOTP are not compact, and some non-interactive ones—the CL scheme for Clifford circuits and the AUX scheme for circuits with constant $T$-depth—provided by Broadbent and Jeffery[11] are compact. Subsequently, Dulek and Schaffner extended Broadbent's work and proposed a QHE scheme for polynomial-sized $T$ gates utilizing another kind of $T$ gadget[12]. In addition, other issues, such as multi-party decryption, circuit privacy, quasi-compactness, and generalization to classical encrypted data are also investigated in various QHE schemes[13–15]. It should be noted that introducing CHE (combining with some sophisticated techniques) into QHE is double edged: it not only reduces the communication complexity and gains compactness or quasi-compactness but also compromises security and narrows the schemes' applicability. These hybrid QHE schemes cannot reach perfect security as many pure quantum cryptosystems do, and they become inefficient for circuits with a large number of $T$ gates (e.g. the circuit of Grover's search).

[1]School of Computer Science and Technology, Huazhong University of Science and Technology, Wuhan, 430074, China. [2]Hubei Engineering Research Center on Big Data Security, School of Cyber Science & Engineering, Huazhong University of Science and Technology, Wuhan, 430074, China. [3]Shenzhen Huazhong University of Science and Technology Research Institute, Shenzhen, 518063, China. [4]College of Mathematics and Statistics, Shenzhen University, Shenzhen, 518060, China. *email: lusongfeng@hotmail.com; yqcui1977@hust.edu.cn

| Variables and notations | Explanations |
|---|---|
| $M, m, n(m, n \in \mathbb{N}^+)$ | $M = 2^m$ is the number of items to be searched, and each item data($j$) ($j < M$) contains $n$ bits. |
| $ek, dk, sk, eek, ddk$ | $ek$ and $dk$ are encryption and decryption keys for Alice's data, respectively; $sk$ is used to encrypt $dk$ in Protocol 1; $eek$ and $ddk$ are used to encrypt $ek$ and recover $dk$, respectively, in Protocol 2. |
| $x_r(k), z_r(k)(r, k \in \mathbb{N})eek(k)$ | $(x_r, z_r)$ $(r > 0)$ is the $2n$-bit intermediate key for the $r^{\text{th}}$ round of key-updating; $x_0$ and $z_0$ constitute $ek$; $x_r(k)$, $z_r(k)$, and $eek(k)$ are the $k^{\text{th}}$ bit of $x_r$, $z_r$, and $eek$, respectively. |
| $X_i, Y_i, Z_i, H_i, S_i, T_i\text{CNOT}_{i,l}$ | $X_i, Y_i, Z_i, H_i, S_i$ or $T_i$ denotes applying a $X$, $Y$, $Z$, $H$, $S$ or $T$ gate on the $i^{\text{th}}$ qubit of the input state and letting the other qubits unchanged; $\text{CNOT}_{i,l}$ denotes performing a CNOT gate on the $i^{\text{th}}$ and $l^{\text{th}}$ qubits of the input, which act as the control and target qubits, respectively. Note that $S$ and $T$ gates are called $P$ and $R$ gates, respectively, in ref. [8]. |

**Table 1.** Explanations for frequently used variables and notations.

Apart from the foregoing QHE schemes based on CHE and QOTP, constructions utilizing other techniques are explored, such as a private-key QHE protocol based on the centralizer of a subgroup of operations[16], a practical somewhat-secure QHE on coherent states[17], a QHE scheme for circuits with constant number of $T$ gates based on quantum codes[18], and a QHE scheme with multiple evaluators based on $(k,n)$-threshold quantum state sharing[19]. These schemes only apply to limited circuit families.

In consideration of the huge number of iterations in Grover's search, where each Grover iteration contains $T$ gates, we need to fall back on the early QHE schemes with interactions[6,8,9] rather than non-interactive ones[11,12] based on CHE, because the latter is unable to handle so many $T$ gates. To reduce Alice's workload, a trusted third party is introduced. To prevent Bob from knowing the original search condition (say, $C_{\text{o}}$), the proposed search protocol encrypts the qubits involved in $C_{\text{o}}$ twice and let Bob searches on the cipher state according to a new search condition $C_{\text{n}}$, which can be viewed as the cipher-text of $C_{\text{o}}$.

As for Clifford circuits without any $T$ gate, their homomorphic evaluations become much easier, since the Clifford gates can be applied on a cipher state without causing any undesirable error. For this type of circuits, a compact and perfectly secure homomorphic evaluating protocol is proposed. A main distinction between the two protocols lies in the different treatments on the key-update process, which is an important part of QOTP-based QHE.

## Key Update in Quantum Homomorphic Encryption Based on Quantum One-time Pad

A QHE scheme consists of four components: key generation, encryption, decryption, and homomorphic evaluation. In a QOTP-based QHE scheme, quantum states are encrypted or decrypted using QOTP, which transforms the input state qubit by qubit with Pauli operators $\{X, Z\}$ depending on a classical encryption key $ek = (x_0, z_0)$. In this scheme, the homomorphic evaluation for a Clifford gate is to apply this gate directly on an encrypted state, whereas the evaluation for a non-Clifford gate (the $T$ gate) is to perform a $T$ gadget on the cipher-state. Additionally, the decryption key for $ek$ needs to be refreshed synchronously with respect to the evaluation.

For clarity, we rephrase the key-update rules for arbitrary unitary transforms and quantum measurements in Algorithms 1 and 2, respectively, which are derived from refs. [7,8]. The frequently used variables in this paper are listed in Table 1.

**Algorithm 1** (Key update for unitary transforms). Suppose that $|\psi\rangle$ is an $n$-qubit state, $V$ is an $n$-qubit unitary transform consisting of gates from the universal gate set $\mathcal{G} = \{I, X, Y, Z, H, S, \text{ CNOT }, T\}$, and $G$ is a two-level unitary transform that merely performs one gate from $\mathcal{G}$ on two or fewer qubits. Let $ek = (x_0, z_0)$ be an encryption key and $dk_r = (x_r, z_r)$ the decryption key for $V$ and $ek$, such that $V(\otimes_{k=1}^{n} Z^{z_0(k)} X^{x_0(k)})\ |\psi\rangle = (\otimes_{k=1}^{n} Z^{z_r(k)} X^{x_r(k)})\, V\ |\psi\rangle$. Then the updated decryption key $dk_{r+1} = (x_{r+1}, z_{r+1})$ for $G \cdot V$ and $ek$ satisfying $G \cdot V(\otimes_{k=1}^{n} Z^{z_0(k)} X^{x_0(k)})|\psi\rangle = (\otimes_{k=1}^{n} Z^{z_{r+1}(k)} X^{x_{r+1}(k)}) G \cdot V|\psi\rangle$ is calculated as follows:

- If $G = I, X_i, Y_i, \text{or} Z_i$, then $dk_{r+1} = dk_r$.
- If $G = H_i$, then $(x_{r+1}(i), z_{r+1}(i)) = (z_r(i), x_r(i))$.
- If $G = S_i$, then $(x_{r+1}(i), z_{r+1}(i)) = (x_r(i), x_r(i) \oplus z_r(i))$.
- If $G = \text{CNOT}_{i,l}$, then $(x_{r+1}(i), z_{r+1}(i)) = (x_r(i), z_r(i) \oplus z_r(l))$ and $(x_{r+1}(l), z_{r+1}(l)) = (x_r(i) \oplus x_r(l), z_r(l))$.
- If $G = T_i$ (suppose the secret bits Alice chooses for this $T$ gate are $\hat{y}$ and $\hat{d}$, and the related one-bit measurement result from the server is $\hat{c}$), then $(x_{r+1}(i), z_{r+1}(i)) = [x_r(i) \oplus \hat{c}, x_r(i) \cdot (\hat{c} \oplus \hat{y} \oplus 1) \oplus z_r(i) \oplus \hat{d} \oplus \hat{y}]$ (where $\oplus$ denotes addition modulo 2).

**Algorithm 2** (Key update for quantum measurements). Let $|\widetilde{\psi}\rangle = (\otimes_{k=1}^{n} Z^{z_r(k)} X^{x_r(k)})\ |\psi\rangle$ be an $n$-qubit cipher state under key $dk_r = (x_r, z_r)$ and $\mathscr{M}$ a computational basis measurement. Then the updated decryption key $dk_{r+1}$ for the classical measurement outcome $\mathscr{M}\ |\widetilde{\psi}\rangle \equiv \widetilde{\psi}_{\mathscr{M}}$ satisfying $\widetilde{\psi}_{\mathscr{M}} \oplus dk_{r+1} = \psi_{\mathscr{M}}$ is $dk_{r+1} = x_r$, where $\psi_{\mathscr{M}}$ is the underlying plaintext of $\widetilde{\psi}_{\mathscr{M}}$.

It follows that, the final decryption key $dk$ for a unitary transform $U$ depends on the encryption key $ek$, the concrete circuit $\mathscr{C}_U$ of $U$, the secret parameters $(y, d)$ for the $T$ gates in $\mathscr{C}_U$, and the measurement result $c$ corresponding to those $T$ gates. Hence, during a homomorphic execution of $\mathscr{C}_U$, there is a unitary decryption-key-obtaining transform $DK_{\mathscr{C}_U}$ that satisfies the equation $|ek, y, d, c, dk\rangle = DK_{\mathscr{C}_U}\ |ek, y, d, c, 0\rangle$, where $|0\rangle$ serve as the auxiliary qubits in workspace.

A critical barrier on the route to an efficient and perfectly secure quantum fully homomorphic encryption scheme is caused by the $T$ gate, which, however, is an indispensable ingredient of universal gate set. To homomorphically perform a $T$ gate, a phase correction depending on the encryption key has to be applied, whereas the key cannot be revealed to the server. To overcome this obstacle, some schemes introduce a round of communication conveying some evaluated information determined by the encryption key between the client and the server[6,8,9], whereas others resort to classical homomorphic encryption[11,12].

On the other hand, homomorphically performing a Clifford circuit (not involving any $T$ gate) could be more convenient. Before we present this, we conclude an important feature of the key-update procedure for Clifford circuits in Lemma 1, which will be used in our second protocol.

**Lemma 1**. The key-update process corresponding to a Clifford circuit $\mathscr{C}$ acting on $n$ qubits can be encoded as a $4n^2$-bit string $str_{\text{ku}}^{\mathscr{C}}$, whose computational complexity is $O(n\log 2n)$.

*Proof.* According to Algorithm 1, the key-update operations for gates except $T$ only involve bit exchanges and some $\oplus$. Since bit exchange $(d_j, d_k) \to (d_k, d_j)$ (where $d_j$ is the $j^{\text{th}}$ bit of $d$, $j \neq k$, and $d \in \{0, 1\}^{2n}$) can be implemented with three $\oplus: (d_j, d_k) \to (d_j, d_j \oplus d_k), (d_j, d_j \oplus d_k) \to (d_j \oplus (d_j \oplus d_k), d_j \oplus d_k) = (d_k, d_j \oplus d_k), (d_k, d_j \oplus d_k) \to (d_k, (d_j \oplus d_k) \oplus d_k) = (d_k, d_j)$, the key-update procedure for $\mathscr{C}$ can be implemented by a sequence of commands of the form $(d_j, d_k) \to (d_j, d_j \oplus d_k)$; that is, add (modulo 2) one key bit $d_j$ to a target key bit $d_k$. During this procedure, each key bit may act as a target and other key bits that are *effectively* added to this target constitute a set of size less than $2n$. Here, a bit $d_j$ is *effectively* to $d_k$ if $d_j$ is added to $d_k$ odd times. Note that the operations that adding one bit to another bit for even times are also part of the key-update process and convey considerable information about $\mathscr{C}$, which are ignored due to the calculating rule of $\oplus$. Thus, each decryption key bit can be expressed as $u_0 \cdot d_0 \oplus u_1 \cdot d_1 \oplus \cdots \oplus u_{2n-1} \cdot d_{2n-1}$ $(u_j \in \{0, 1\})$, which can be encoded as a $2n$-bit string $u_0 u_1 \ldots u_{2n-1}$. Since $2n$ additions can be parallelized, they can be computed with $\log 2n$ steps. As a result, the complete decryption key can be encoded as a $4n^2$-bit string $str_{\text{ku}}^{\mathscr{C}}$ and be computed with $O(n\log 2n)$ steps. □

## Quantum Search on Encrypted Data

Since Grover's algorithm excepting the final measurement is a unitary transform, one can certainly apply it homomorphically on a cipher-state and then obtain the encrypted search result. Here, we consider the situation that Alice wants Bob to search on her encrypted superposition state (which could be obtained by a QRAM addressing scheme[20]) but does not care how he carries out the operations exactly. In other words, Alice prefers to prepare the input data and handles the evaluated output rather than participate in the entire process of the evaluation. This requirement is reasonable in the sense that it coincides with the relationship between the client and the server in cloud computing, which is a prevalent paradigm nowadays.

It is known that Grover's search is made up of a sequence of Grover iterations, each of which contains an oracle that is able to mark items satisfying a specific search condition. When it is used to speed up the solution of an NP problem of size $m$ (there are at least $2^m$ candidates for the solution of this problem), the oracle requires $O(\text{poly}(m))$ elementary gates (where $\text{poly}(m)$ is a polynomial in $m$), since a solution of an NP problem can be recognized in polynomial time. Since each Grover iteration contains $\text{poly}(m)\,T$ gates, the total number of $T$ gates in the search circuit reaches $O(2^{m/2} \cdot \text{poly}(m))$, which is beyond the capability of CHE-based QHE schemes. A dilemma seems to appear: the key shall neither be updated homomorphically using CHE by Bob[11,12] nor be renewed interactively by Alice. As a way out of this impasse, we introduce a third party—a trusted key center, Carol—between Alice and Bob, to undertake the interactive work.

**Outsourcing key update to a trusted key center.** In QHE, key generation includes randomly choosing a classical encryption key and calculating the decryption key using the key update algorithm, which becomes long and tedious if there are a large number of $T$ gates in the evaluation. To ease the burden of Alice and to keep the homomorphic search going successfully, we introduce a trusted key center, Carol, to negotiate a random encryption key with Alice and then to calculate the corresponding decryption key by communicating with Bob. That is, we divide the client in the interactive QHE scheme[9] into two parts: a thin client (Alice) and a trusted key center (Carol). The requirements and constraints on Carol are given in Constraint 1.

**Constraint 1** (Requirements and constraints on the key center). The key center Carol obeys the following four constraints:

1. Carol is a classical computer augmented with the ability to prepare qubits in $\{|+\rangle = (|0\rangle + |1\rangle)/\sqrt{2}, |-\rangle = (|0\rangle - |1\rangle)/\sqrt{2}, |+_y\rangle = (|0\rangle + i\,|1\rangle)/\sqrt{2}, |-_y\rangle = (|0\rangle - i\,|1\rangle)/\sqrt{2}\}$, which serve as auxiliary qubits for $T$ gates in homomorphic evaluations as well as the encoded results of random bits in quantum key distribution (QKD).

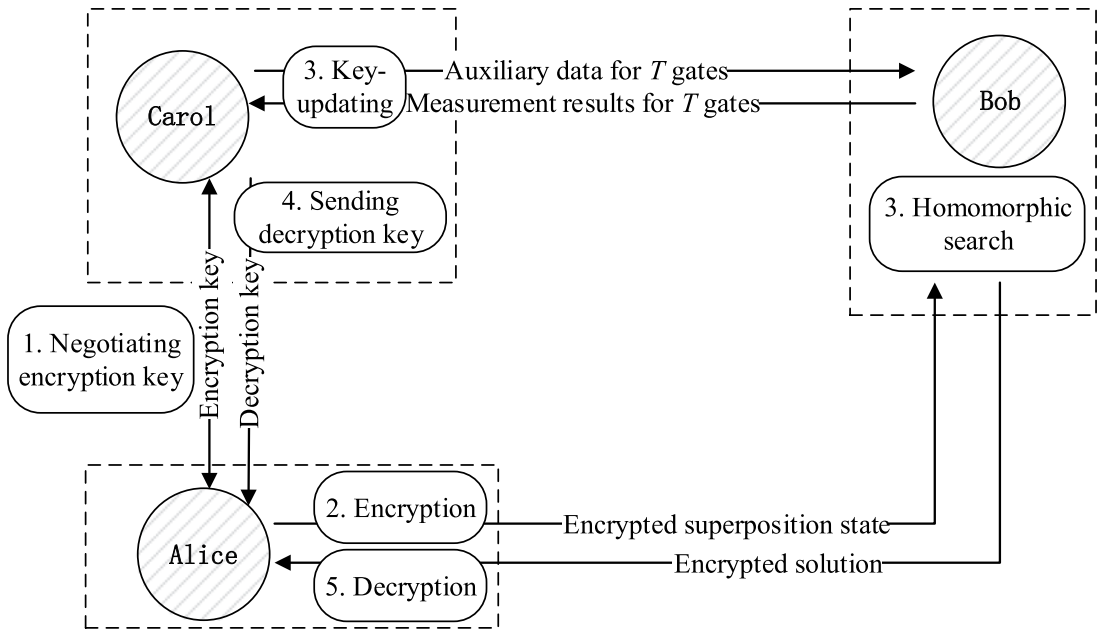

**Figure 1.** Tasks of each party and interactions among parties. First, Alice negotiates an encryption key for her data with Carol. Then, Alice encrypts her superposed data with the encryption key and sends the encrypted state to Bob. Once Bob receives the encrypted data, the homomorphic search and the key-update process (performed by Carol) begin and proceed synchronously. During the homomorphic search, Carol sends some auxiliary data to Bob and then receives some related measurement results, which are needed for key update. When Bob finishes the homomorphic search and obtains an encrypted solution, Carol finishes the key-update process and produces the corresponding decryption key as well. After that, Alice receives the search result and the decryption key from Bob and Carol, respectively. Finally, Alice decrypts the ciphered search result with the decryption key.

2. Carol honestly obeys the subsequent search protocol and would not reveal the encryption and decryption keys to parties other than Alice.
3. Carol would not reveal the circuit detail (including the auxiliary states and the measurement outcomes) performed by Bob to parties other than Bob.
4. Carol would not be corrupted by any adversary who wishes to obtain some information about the plain input provided by Alice or the quantum circuit performed by Bob.She would not reveal the key-update procedure (including the intermediate keys) to others (including Alice and Bob).

Our search protocol runs among Alice (the client), Bob (the search server), and Carol (the trusted key center) as illustrated in Fig. 1, where key generation is split into three parts: Alice negotiates an encryption key with Carol, Carol calculates the corresponding decryption key using the key-update algorithm, and Carol securely sends the decryption key to Alice. With the help of QKD and classical one-time pad (COTP), these tasks can be accomplished with perfect security. Moreover, since the key update operations are classical, Carol could be a classical computer augmented with the ability to prepare only four types of quantum states listed in Constraint 1. In this protocol, Alice needs to prepare superposed data items using QRAMs and to perform single-qubit measurements (in the Pauli-$X$ and $Y$ bases), Pauli operators $\{X, Z\}$, and classical bitwise additions (modulo 2) to implement the BB84 protocol, encrypt the plain data, and decrypt the evaluated result, respectively.

**A homomorphic search protocol with a key center.** With the assistance of the key center, an interactive protocol for quantum search on encrypted data proceeds as shown in Protocol 1. For clarity, we call the the collection of qubits (identified by their locations in Alice's data $|\psi_0\rangle$) appearing in the plain (original) search condition $C_o$ "the search field" of $|\psi_0\rangle$, and denote it by $|\zeta\rangle$, which is a part of $|\psi_0\rangle$. In the first two steps, Alice encrypts $|\zeta\rangle$ and produces a modified search condition to conceal the original one from Bob, so that Bob knows nothing about her plain search target. To briefly explain this, suppose that $|\zeta\rangle$ consists of the 3rd, 5th and 6th qubits of $|\psi_0\rangle$, Alice wants to find out the items satisfying the clause $C_o = d_3 \vee \overline{d_5} \vee d_6$ (a clause in 3-conjunctive normal form, where $d_j$ is the $j$th bit of a data item $d$, $\overline{d_j} \equiv 1 \oplus d_j$, the symbol "$\vee$" indicates "OR"), and the 3-bit encryption key $srk$ for $|\zeta\rangle$ is $r_0r_1r_2$ ($r_0, r_1, r_2 \in \{0, 1\}$), then $C_n = (d_3 \oplus r_0) \vee (\overline{d_5} \oplus r_1) \vee (d_6 \oplus r_2)$. From Bob's point of view, $srk$ can be arbitrary 3-bit string, each with the same probability 1/8 and maps the received $C_n$ to one of the eight possible clauses associated with $|\zeta\rangle$, that is, $P(C_o|C_n) = P(C_o)$ for any $C_n$ and any $C_o$, so Bob knows nothing about the plain search condition from a given $C_n$.

The schematic circuit for this protocol is depicted in Fig. 2. To make our circuit more comprehensible, we adapt the $T$ gadget in ref. [8] to Fig. 2 and illustrate the $j^{th}$ $T$ gadget in the $l^{th}$ $G$ iteration on the $g^{th}$ wire in Fig. 3, where $|\psi_g\rangle$ is the $g^{th}$ qubit of the plain state $|\psi\rangle$, and $Z^{\hat{z}}X^{\hat{x}}\ |\psi_g\rangle$ is the encrypted qubit of $|\psi_g\rangle$ just before the application of this $T$ gadget.

**Protocol 1**. Homomorphic quantum search on encrypted states

1. Alice prepares a plain superposition state $|\psi_0\rangle = \frac{1}{\sqrt{M}}\sum_{j=0}^{M-1}|\ \text{data}\ (j)\rangle$ from her classical database.
2. Alice encrypts the search fields $|\zeta\rangle$ (within her plain state $|\psi_0\rangle$) with a random encryption key $srk$ using QOTP and obtains an encrypted result $Enc_{srk}\ |\zeta\rangle$, which is a part of the partially encrypted state $|\psi\rangle = Enc_{srk}\ |\psi_0\rangle$ (Here, $Enk_{srk}$ is the extended encryption transform on $|\psi_0\rangle$, which applies the encryption on $|\zeta\rangle$ and leaves the other qubits unaffected).
3. Alice translates the original search condition $C_o$ (on $|\zeta\rangle$) into the new condition $C_n$ on $Enc_{srk}\ |\zeta\rangle$ such that $|\zeta\rangle$ satisfies $C_o$ if and only if $Enc_{srk}\ |\zeta\rangle$ satisfies $C_n$.

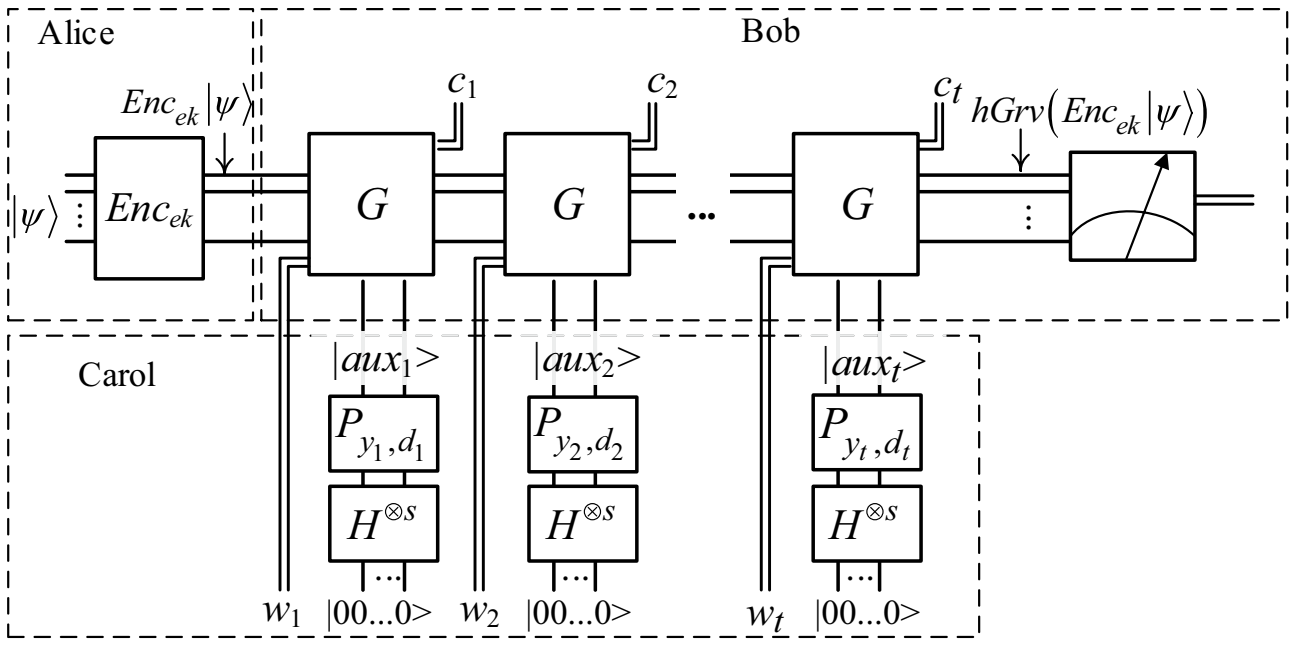

**Figure 2.** Schematic circuit implementing Protocol 1. There are $t = O(\sqrt{M})$ $G$ iterations in the circuit, which carries out the homomorphic search transform denoted by $hGrv$. A $G$ iteration is a homomorphic version of the standard Grover iteration, which is obtained by replacing each $T$ gate in the normal Grover iteration with the $T$ gadget illustrated in Fig. 3. Suppose there are $sT$ gadgets in each $G$ iteration; then, the $l^{th}$ $G$ iteration ($l \in \{1, 2, \ldots, t\}$) needs an auxiliary state $|aux_l\rangle$ of $s$ qubits and an evaluation key $w_l$ of $s$ bits to perform $sS$ corrections. $P_{y_l,d_l} \equiv \otimes_{j=1}^{s} P_{y_l(j),d_l(j)}$ is the phase transform for the $l^{th}$ $G$ iteration, where $y_l$ and $d_l$ are $s$-bit binary strings randomly chosen by Carol for all $T$ gadgets in the $l^{th}$ $G$ iteration, and $P_{y_l(j),d_l(j)} \equiv Z^{d_l(j)}S^{y_l(j)}$ is the phase transform for the $j^{th}$ $T$ gadget in the $l^{th}$ $G$ iteration. Applying $P_{y_l(j),d_l(j)}$ on $|+\rangle$ gives the $j^{th}$ qubit of $|aux_l\rangle$, which is $|+\rangle$, $|-\rangle$, $|+_y\rangle$ or $|-_y\rangle$. $w_l$ is the Boolean XOR result between some intermediate key bits and $y_l$; $c_l$ is an $s$-bit measurement result from the $l^{th}$ $G$ iteration. $y_l$, $d_l$ and $c_l$ are involved in key update for the $l^{th}$ $G$ iteration.

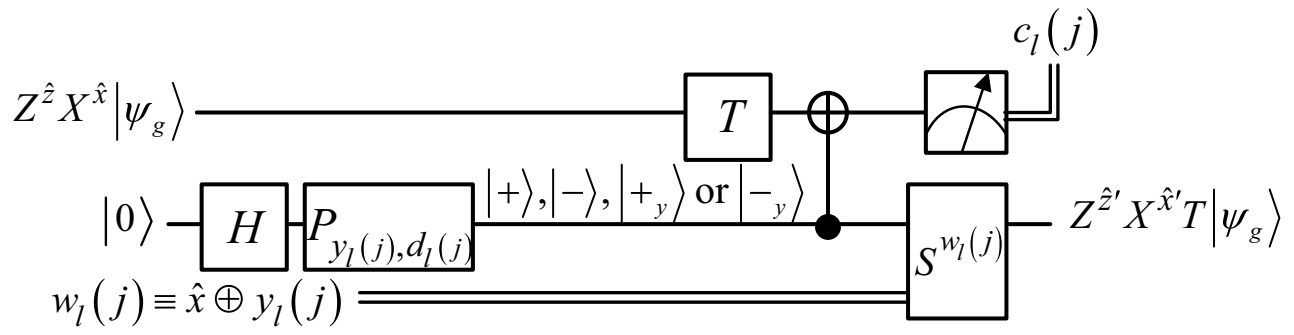

**Figure 3.** The $T$ gadget within the schematic circuit.

4. Alice sends a number $n$—the length of her encrypted state—to Carol.
5. Carol shares a binary string of $3n$ random bits with Alice by the BB84 protocol[21], where $|+\rangle$, $|+_y\rangle$ stands for 0, and $|-\rangle$, $|-_y\rangle$ stands for 1. The first $2n$ bits of the binary string act as $ek$ and the remaining $n$ bits act as the encryption key $sk$ for $dk$.
6. Alice encrypts $|\psi\rangle$ with $ek = (x_0, z_0)$ and sends the encrypted state $Enc_{ek}|\psi\rangle = \otimes_{k=1}^{n} Z^{z_0(k)}X^{x_0(k)}|\psi\rangle$ along with $C_n$ to Bob.
7. Bob searches on $Enc_{ek}|\psi\rangle$ homomorphically using $hGrv$, and Carol updates the key synchronously. During the search, once a $T$ gate appears, Bob asks Carol to send an auxiliary qubit from $\{|+\rangle, |-\rangle, |+_y\rangle, |-_y\rangle\}$ along with a related evaluating key bit ($w_l(j)$ in Fig. 3) to him and then performs the $T$ gadget with those data. After that, he returns a one-bit measurement result ($c_l(j)$ in Fig. 3) to Carol for key update.
8. When the homomorphic search is completed, Bob measures the search result $hGrv(Enc_{ek}|\psi\rangle)$ and returns the (encrypted) classical outcome $\mathcal{M}[hGrv(Enc_{ek}|\psi\rangle)]$ to Alice. Meanwhile, Carol finishes the key update process and obtains the decryption key $dk$ for $ek$ and $hGrv$; she then encrypts $dk$ with $sk$ using COTP and sends it to Alice.
9. Alice first recovers $dk$ with $sk$ and decrypts $\mathcal{M}[hGrv(Enc_{ek}|\psi\rangle)]$ with $dk$ to obtain the partially encrypted search result, then recovers the search fields in the search result with $srk$. After that, she can check this result with her search condition and decide whether Bob honestly performed the homomorphic search.

In step 1, Alice needs to utilize two kinds of QRAMs: the first one maps $\frac{1}{\sqrt{M}}\sum_{j=0}^{M-1}|j\rangle\,|0\rangle$ to $\frac{1}{\sqrt{M}}\sum_{j=0}^{M-1}|j\rangle\,|\text{data}(j)\rangle$, which is connected to a memory array where data ($j$) is stored in the $j$th memory cell; the second one maps $\frac{1}{\sqrt{M}}\sum_{j=0}^{M-1}|j\rangle\,|\text{data}(j)\rangle$ to $\frac{1}{\sqrt{M}}\sum_{j=0}^{M-1}|0\rangle\,|\text{data}(j)\rangle$, which is connected to a memory where $j$ is stored in the data($j$)th memory cell (we assume that data($i$) $\neq$ data($j$) if $i \neq j$).

In Fig. 2, Carol gives $|aux_l\rangle = P_{y_l,d_l} \otimes_{j=1}^{s} H|0\rangle$ and $w_l$ to Bob for the $l^{th}$ $G$ iteration, and Bob then returns an $s$-bit measurement result $c_l$ from this iteration to Carol, which is used to update the key. More precisely, as illustrated in Fig. 3, the $j^{th}$ qubit of $|aux_l\rangle$ and the $j^{th}$ bit of $w_l$ (or $w_l(j)$) together with an encrypted qubit $Z^{\hat{z}}X^{\hat{x}}|\psi_g\rangle$ under key $(\hat{x}, \hat{z})(\hat{x}, \hat{z} \in \{0, 1\})$ act as the input of the $j^{th}$ $T$ gadget in the $l^{th}$ $G$ iteration. This $T$ gadget then outputs a one-bit measurement outcome $c_l(j)$ and an encrypted result of $T|\psi_g\rangle$ under a refreshed key $(\hat{x}', \hat{z}')$ $(\hat{x}', \hat{z}' \in \{0, 1\})$, which satisfies $\hat{x}' = \hat{x} \oplus c_l(j)$ and $\hat{z}' = \hat{x}[c_l(j) \oplus y_l(j) \oplus 1] \oplus \hat{z} \oplus d_l(j) \oplus y_l(j)$. The classical decryption (performing a bitwise addition between the encrypted search result and the decryption key) after the quantum measurement is omitted.

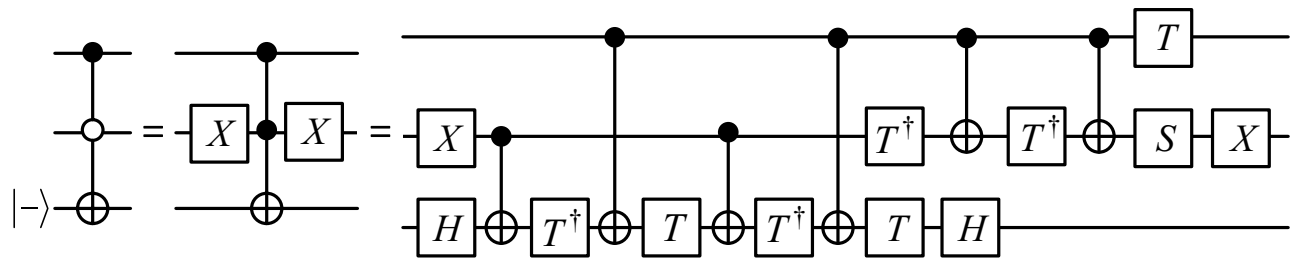

**Figure 4.** The oracle for finding 10 out of {00, 01, 10, 11}.

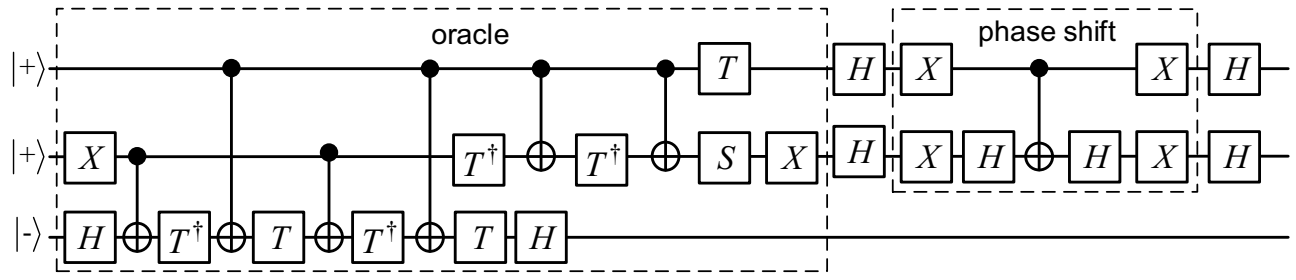

**Figure 5.** The two-qubit Grover's search circuit.

**Simulation of two-qubit homomorphic search.** To verify the effectiveness of Protocol 1, we present the detailed circuit for two-qubit homomorphic search in Supplementary Fig. S1 and simulate its process using MATLAB. The detailed simulation code is listed in the Supplementary File, which always outputs the expected search target.

Without loss of generality, suppose Alice wants to find the item 10 out of the set {00, 01, 10, 11} (the partially encrypted results corresponding to $Enc_{srk}\ |\psi_0\rangle$); thus, a single Grover iteration is enough, and the oracle within the iteration can be built with $X$ and Toffoli gates as illustrated in Fig. 4[22]. After decomposing the Toffoli gate into the universal gates supporting homomorphic evaluations (i.e. $X$, $Z$, $H$, $S$, CNOT, $T$), a quantum circuit for two-qubit plain-state search is obtained as depicted in Fig. 5. To perform a two-qubit cipher-state search, this circuit needs to be transformed into its homomorphic version, which is achieved by replacing each $T$ gate (or $T^\dagger$ gate) in Fig. 5 with a $T$ gadget (or $T^\dagger$ gadget). The $T^\dagger$ gadget is obtained by replacing each $S$ gate in Fig. 3 with an $S^\dagger$ gate. The key-update rules for $T^\dagger$ and $T$ gadgets are identical.

With QOTP encryption and COTP decryption being included, the resultant homomorphic quantum search circuit for two-qubit states is presented in Supplementary Fig. S1. The input of this circuit is the superposed (partially encrypted) state $|\psi\rangle = \frac{1}{2}(|00\rangle + |01\rangle + |10\rangle + |11\rangle)$ along with the oracle qubit $|-\rangle$, and the output should be the classical search target (still partially encrypted) together with the unchanged oracle qubit.

In outline, the simulation proceeds in six steps, where the operations carried out by the three parties are included.

1. Randomly generate a four-bit encryption key $ek = (x_0, z_0)$ for $|\psi\rangle$.
2. Randomly generate a 14-bit evaluation key $evk = (y_1, d_1)$ for the seven $T$ gadgets (in Supplementary Fig. S1).
3. Encrypt $|\psi\rangle\ |-\rangle$ with $ek$ using QOTP. The encrypted result acts as the input state *encState* of the quantum homomorphic search.
4. Perform the homomorphic search circuit on *encState* and obtain the encrypted output state. Meanwhile, refresh the intermediate key using the key update algorithm.
5. Measure the encrypted output state; update the intermediate key and then get a two-bit decryption key $dk$.
6. Decrypt the output cipher-state with $dk$ using COTP and get the search result.

The simulation indicates that with different encryption keys ($ek = (x_0, z_0)$), evaluation keys ($evk = (y_1, d_1)$) and intermediate measurement results ($c_1$), one may get different decryption keys ($dk$) and different encrypted search results, but the decrypted outcome is always 10—the partially encrypted search target that we predetermined at the beginning of this section. Here, we list some representative outcomes in Table 2, where the last bit of $x_0$ and the last bit of $z_0$ stay 0 since the oracle qubit does not need to be encrypted.

## A Compact Quantum Homomorphic Evaluation Protocol for Clifford Circuits

In this section, we demonstrate a non-interactive homomorphic evaluation protocol for Clifford circuits, where the decryption key can be calculated by Alice with polynomial overhead (with respect to the size of $ek$) by Lemma 1.

**Protocol 2**. Quantum homomorphic evaluation for a Clifford circuit $\mathscr{C}$.

1. Alice randomly chooses a $2n$-bit encryption key $ek$ for her $n$-qubit plain input $|\psi\rangle$.
2. Alice encrypts her plain state $|\psi\rangle$ with $ek$ and sends the ciphered output $Enc_{ek}\ |\psi\rangle = \otimes_{k=1}^{n} Z^{z_0(k)} X^{x_0(k)}\ |\psi\rangle$ to Bob.
3. Bob initiate a $4n^2$-bit binary string $str_{\mathrm{ku}}^{\mathscr{C}}$, which will encode the simplified key-update expressions for $\mathscr{C}$, then applies $\mathscr{C}$ on $Enc_{ek}\ |\psi\rangle$. During the performance of $C$, he synchronously updates $str_{\mathrm{ku}}^{\mathscr{C}}$. When the evaluation is completed, Bob returns the evaluated result $U_{\mathscr{C}}(Enc_{ek}\ |\psi\rangle)$ and $str_{\mathrm{ku}}^{\mathscr{C}}$ to Alice.

| input state | $ek$ | $evk$ | $c_1$ | Encrypted result | $dk$ | Decrypted result |
|---|---|---|---|---|---|---|
| $\|+\rangle\|+\rangle$ | $x_0 = 100$ | $y_1 = 1010110$ | 1110111 | 01 | 11 | 10 |
| | $z_0 = 110$ | $d_1 = 0111001$ | | | | |
| | $x_0 = 100$ | $y_1 = 1110011$ | 1000010 | 00 | 10 | |
| | $z_0 = 010$ | $d_1 = 1011010$ | | | | |
| | $x_0 = 100$ | $y_1 = 0100001$ | 0000101 | 01 | 11 | |
| | $z_0 = 100$ | $d_1 = 0000101$ | | | | |
| | $x_0 = 010$ | $y_1 = 0000011$ | 1001101 | 11 | 01 | |
| | $z_0 = 110$ | $d_1 = 0111110$ | | | | |
| | $x_0 = 110$ | $y_1 = 0101100$ | 1011101 | 10 | 00 | |
| | $z_0 = 110$ | $d_1 = 0001010$ | | | | |

**Table 2.** Outcomes of the simulation.

4. Alice performs the simplified key-update process indicated by $str_{\text{ku}}^{\mathscr{C}}$ on $ek$ and obtains the decryption key $dk$, then decrypts $U_{\mathscr{C}}(Enc_{ek}\,|\psi\rangle) = Enc_{dk}(U_{\mathscr{C}}\,|\psi\rangle)$ and gets the plain evaluated result $U_{\mathscr{C}}\,|\psi\rangle$.

In step 3, $U_{\mathscr{C}}$ is the unitary transform corresponding to $\mathscr{C}$. The simplified key-update expressions for $\mathscr{C}$ consists of $2n$ binary polynomials $\{p_j\}_{j=0}^{2n-1}$, where $p_j$ indicates the relationship between $ek$ and $dk_j$. Concretely, the $k^{\text{th}}$ bit of $ek$ is denoted by a binary-coefficient monomial $x^k$ (which satisfies $x^k \oplus x^k = 0$) and $dk_j$ is a binary polynomial $dk_j = u_0x^0 \oplus u_1x^1 \oplus \cdots \oplus u_{2n-1}x^{2n-1}$ ($u_j \in \{0, 1\}$). Since $dk = ek$ before the evaluation, Bob initially sets $p_j = x^j$ for all $j$, which corresponds to the initial value $\{2^0, 2^1, \ldots, 2^{2n-1}\}$ of $str_{\text{ku}}^{\mathscr{C}}$. As an example, suppose the key-update commands for $\mathscr{C}$ are sequentially $(dk_i, dk_j) \to (dk_i, dk_j \oplus dk_i)$, $(dk_j, dk_k) \to (dk_j \oplus dk_k, dk_k)$, $(dk_i, dk_l) \to (dk_i \oplus dk_l, dk_l)$, $(dk_i, dk_j) \to (dk_i \oplus dk_j, dk_j)$, then Bob updates the corresponding polynomials as follows: (1) $p_j = p_j \oplus p_i = x^j \oplus x^i$; (2) $p_j = p_j \oplus p_k = (x^j \oplus x^i) \oplus x^k$; (3) $p_i = p_i \oplus p_l = x^i \oplus x^l$; (4) $p_i = p_i \oplus p_j = (x^i \oplus x^l) \oplus (x^j \oplus x^i \oplus x^k) = x^j \oplus x^k \oplus x^l$. At this point, the $i^{\text{th}}$ and $j^{\text{th}}$ substrings of $str_{\text{ku}}^{\mathscr{C}}$ become $2^j + 2^k + 2^l$ and $2^i + 2^j + 2^k$, respectively.

## Security and Performance

In what follows, we discuss four important properties—data security, circuit privacy, computational complexity, and compactness—of Protocol 1 and Protocol 2, which characterize the security level and performance of each protocol.

**Data security and circuit privacy.** Since protocol 1 involves a trusted third party (TTP), which does not appear in standard quantum homomorphic encryption schemes, we thus define a new type of security for quantum homomorphic evaluation protocols assisted by TTP. Intuitively, such type of security can be achieved by a quantum cryptographic protocol with a well-defined TTP if the presence of this TTP can preserve the same level of security as that of the underlying (perfectly secure) cryptographic primitives (without TTP).

**Definition 1**. (TTP-assisted perfect security). Let $\Pi$ be a quantum homomorphic evaluation protocol assisted by a trusted third party, Carol, who honestly follows $\Pi$ and would not be corrupted by any adversary. For an arbitrary plain-state $\rho_M$ (the input state, a density matrix) provided by Alice, if

1. The cipher-states of $\rho_M$ (including the initial cipher-state that has not yet been evaluated, the intermediate cipher-states that have undergone some unitary transforms, and the output cipher-state of the evaluation, which can be accessed by the evaluator Bob) are all equivalent to the totally mixed state;
2. The decryption keys for the input state and the ultimate output (a quantum state or a classical binary string after quantum measurement) cannot be (undetectably) obtained by parties other than Alice and Carol; and
3. The decryption keys for the intermediate outcomes during the homomorphic evaluation cannot be obtained by parties other than Carol;

then $\Pi$ satisfies TTP-assisted perfect security.

The first condition in the above is due to the definition of perfect security for quantum encryption schemes[23,24]. A requirement for the (classical) evaluated result after quantum measurement is not included in Definition 1, since the measured outcome of a totally mixed state would not reveal any information about $\rho_M$: the probability of getting a specific classical ciphertext $C$ ($C \in \{0, 1\}^n$) is $\text{tr}\,[|C\rangle\langle C|C\rangle\langle C|(I/2^n)] = 1/2^n$, namely the a posteriori distribution of the ciphertext given $\rho_M$ is the same as the a priori distribution of a valid cryptogram of size $n$.

Definition 1 is mainly concerned with data privacy, which ensures that the plain-states of the input, the intermediate outcomes, and the output can only be accessible to legitimate parties. Taking into account circuit privacy, we exclude Alice from knowing the intermediate keys (calculated by Carol) in the last condition. In particular, if Alice has two intermediate keys $(x_r, z_r)$ and $(x_{r+1}, z_{r+1})$ for the cipher-states immediately before and after an elementary Clifford gate, then she can deduce which kind of gate was performed by comparing $(x_r, z_r)$ and $(x_{r+1}, z_{r+1})$. With the previous definition, we can now prove that the proposed homomorphic search protocol satisfies this new kind of security.

**Theorem 1**. Protocol 1 satisfies TTP-assisted perfect security.

*Proof.* In Protocol 1, the initial cipher-state $\rho_C^0$ of the plain input $\rho_M$ is an encryption of $\rho_M$ using quantum one-time pad under key $(x_0, z_0)$, where $x_0, z_0 \in \{0, 1\}^n$ and each candidate value is assumed to be chosen with the same probability $1/2^{2n}$. Taking into account all possible key values, the initial cipher-state is

$$\rho_C^0 = \frac{1}{2^{2n}} \sum_{x_0,z_0=0}^{2^n-1} (\otimes_{k=1}^n Z^{z_0(k)} X^{x_0(k)})\ \rho_M\ (\otimes_{k=1}^n X^{x_0(k)} Z^{z_0(k)}) = \frac{1}{2^{2n}} \sum_{x_0,z_0=0}^{2^n-1} Z^{z_0} X^{x_0} \rho_M X^{x_0} Z^{z_0} = \frac{I}{2^n},$$

where $X^{x_0} = \otimes_{k=1}^n X^{x_0(k)}$ and $Z^{z_0} = \otimes_{k=1}^n Z^{z_0(k)}$. The detailed derivation of the equation above can be found in ref. [24]. Thus, $\rho_C^0$ is the totally mixed state. After Bob performs a Clifford gate or a $T$ gadget (denoted by $G_1$) on $\rho_C^0$, it becomes the encrypted result of $G_1 \rho_M G_1^\dagger$ under the updated key $(x_1, z_1)$. According to Algorithm 1, there is a one-to-one correspondence between $(x_0, z_0)$ and $(x_1, z_1)$, which means that the distribution of an original key is the same as that of its updated key. So the intermediate cipher-state after the first elementary gate can be written as

$$\rho_C^1 = \frac{1}{2^{2n}} \sum_{x_1,z_1=1}^{2^n-1} Z^{z_1} X^{x_1} (G_1 \rho_M G_1^\dagger) X^{x_1} Z^{z_1}.$$

By ref. [24],

$$\frac{1}{2^{2n}} \sum_{\gamma,\delta=0}^{2^n-1} X^\gamma Z^\delta \rho Z^\delta X^\gamma = \frac{1}{2^{2n}} \sum_{\gamma,\delta=0}^{2^n-1} Z^\delta X^\gamma \rho X^\gamma Z^\delta = \frac{I}{2^n}$$

for an arbitrary density operator $\rho$. Take $\rho = G_1 \rho_M G_1^\dagger$, $x_1 = \gamma$, and $z_1 = \delta$, the two equations above gives $\rho_C^1 = I/2^n$. Analogously, during the sequential performances of the elementary gates within the homomorphic search circuit, all intermediate cipher-states and the ultimate cipher-state obtained by Bob equal the totally mixed state. Therefore, the first requirement of Definition 1 is satisfied.

The decryption key $ek$ for $\rho_M$ is shared between Alice and Carol by using the BB84 protocol; thus, $ek$ cannot be obtained by any eavesdropper without being detected. Since the decryption key $dk$ for the measurement outcome is encrypted with $sk$ using perfectly secure COTP and $sk$ is also protected under the BB84 protocol, so $dk$ cannot be obtained without $sk$, which cannot be undetectably accessed by parties except Alice and Carol. Therefore, the second requirement of Definition 1 is fulfilled.

Let $\rho_C^j$ $(j \geq 1)$ be the intermediate outcome after the first $j$ elementary gates are applied on $\rho_c^0$ and suppose the total number of quantum gates is $\eta$. To get the decryption key $dk_j$ for $\rho_C^j$, either the value of the decryption key $dk_i$ for $\rho_C^i$ $(0 \leq i < j)$ plus the key-update procedure between $dk_i$ and $dk_j$ (or the gate-sequence between $\rho_C^i$ and $\rho_C^j$) need to be known, or the value of $dk_k$ for $\rho_C^k$ $(j < k \leq \eta)$ plus the key-update procedure between $dk_j$ and $dk_k$ (or the gate-sequence between $\rho_C^j$ and $\rho_C^k$) should be known. According to the third constraint on the key center, the only party except Carol knowing the gate-sequence between $\rho_C^i$ and $\rho_C^j$ or between $\rho_C^j$ and $\rho_C^k$ is Bob, who can obtain neither $dk_i$ nor $dk_k$ from Carol due to the last constraint. According to the second constraint, the only party except Carol knowing $dk_0$ (or $dk$, the partial content of $dk_\eta$) is Alice, who can learn neither the gate-sequence between $\rho_C^0$ and $\rho_C^j$ (or between $\rho_C^j$ and $\rho_C^\eta$) nor the key-update procedure between $dk_0$ and $dk_j$ (or between $dk_j$ and $dk_\eta$) due to the last two constraints on Carol. Therefore, the last requirement of Definition 1 is satisfied.

Consequently, Protocol 1 provides TTP-assisted perfect security. □

Since the privacy of the plain search condition in Protocol 1 has been addressed before the detailed description of this protocol, the security definition (as well as the security proof) only considers the confidentiality of decryption keys and the relationship between the plain and cipher states, so that Definition 1 is applicable to general quantum homomorphic evaluation protocols with TTP.

In the proof of Theorem 1, the derivation of the fulfilment of the first condition in Definition 1 can also be used to prove that the initial cipher-state, every intermediate cipher-states, and the ultimate cipher-state in Protocol 2 are also equivalent to the totally mixed state, so Protocol 2 satisfies (standard) perfect security defined in ref. [24]. Apart from data security, circuit privacy is also an important property of homomorphic encryption. Here, we define two kinds of circuit privacy and then prove that they are achieved by Protocol 1 and Protocol 2, respectively.

**Definition 2**. (TTP-assisted circuit privacy). A quantum homomorphic evaluation protocol $\Pi$ assisted by a trusted third party, Carol, satisfies TTP-assisted circuit privacy if, for any circuit $\mathscr{C}$ supported by $\Pi$,

1. Alice cannot obtain any information about $\mathscr{C}$ from the homomorphically evaluated result $U_{\mathscr{C}}(Enc_{ek}\ |\psi\rangle)$ on any cipher-state $Enc_{ek}\ |\psi\rangle$ (or the measurement outcome of this result), except for what can be learned from the decrypted output state $U_{\mathscr{C}}\ |\psi\rangle$ (or the decrypted result of the measurement outcome).
2. Alice cannot gain any information about $\mathscr{C}$ from Carol.

**Theorem 2**. Protocol 1 satisfies TTP-assisted circuit privacy.

*Proof.* In protocol 1, Alice asks Bob to perform $hGrv$ for her. By Theorem 1, the density operator of the output state $hGrv(Enc_{ek}\ |\psi\rangle)$ is a totally mixed state. Without decryption, this state reveals nothing about the details of the homomorphic search except the input/output length of $hGrv$, which has been known by Alice. Clearly, the measurement outcome of a totally mixed state would not reveal any circuit detail, too. After decryption, Alice can get the plain search result, which is the only thing she can learn about $hGrv$. Therefore, the first requirement of Definition 1 is satisfied.

| Schemes | Security | Compactness | Interaction during evaluation |
|---|---|---|---|
| L15[6] | perfect | not compact | two-way QI |
| FBS14[8] | perfect | not compact | two-way CI & one-way QI |
| B15[9] | perfect | not compact | two-way CI & one-way QI |
| EPR[11] | q-IND-CPA | $R^2$-quasi-compact | non-interactive |
| AUX[11] | q-IND-CPA | compact | non-interactive |
| TP[12] | q-IND-CPA | compact | non-interactive |
| GT[26] | perfect | $R\log R$-quasi-compact | non-interactive |
| VGT[26] | perfect | $R$-quasi-compact | non-interactive |
| Protocol 1 | perfect | not compact | two-way CI & one-way QI |
| CL[11] | q-IND-CPA | compact | non-interactive |
| Protocol 2 | perfect | compact | non-interactive |

**Table 3.** Comparison with related works. The former nine proposals are for circuits containing $T$ gates and the latter two are for Clifford circuits.

According to Algorithm 1, one can derive some details of $hGrv$ if he learns some key-update steps, some intermediate keys, some auxiliary qubits for $T$ gadgets, or some measurement outcomes from $T$ gadgets. However, the last two constraints on the key center ensure that no one can obtain such information from Carol, including Alice. Therefore, the second requirement of Definition 1 is satisfied. □

Due to the deterministic relationship between the key-update process and a Clifford circuit evaluated in a QOTP-based QHE, the circuit information cannot be completely hidden from the key-updater. For schemes using CHE, the key is homomorphically updated by the evaluator, so that circuit privacy can be achieved if a circuit private CHE is adopted[12]. However, the use of CHE leads to q-IND-CPA security rather than perfect security (on input data). Besides, the data owner can always learn $O(n)$ bits of information about the evaluated circuit by looking at the output[14], where $n$ is the input/output size of the circuit. Although a QOTP-based QHE protocol not using CHE cannot achieve standard circuit privacy (where the key is updated by the data owner), it may satisfy bounded circuit privacy, which is a weaker notion for circuit privacy.

**Definition 3.** ($O(n^2)$-bounded circuit privacy). A quantum homomorphic evaluation protocol $\Pi$ with an input of size $n$ satisfies $O(n^2)$-bounded circuit privacy if, for any circuit $\mathscr{C}$ supported by $\Pi$, the information about $\mathscr{C}$ obtained by the data owner is bounded by $O(n^2)$.

**Theorem 3.** Protocol 2 satisfies $O(n^2)$-bounded circuit privacy.

*Proof.* In addition to the evaluated cipher-state $U_{\mathscr{C}}(Enc_{ek}\,|\psi\rangle)$, the data owner Alice gets a $4n^2$-bit string $str_{\rm ku}^{\mathscr{C}}$ that depends on the key-update procedure for $\mathscr{C}$. If the decrypted result $U_{\mathscr{C}}\,|\psi\rangle$ is also included (which surely depends on $\mathscr{C}$ and may be related with $str_{\rm ku}^{\mathscr{C}}$), then the information (after measurement) obtained by Alice does not exceed $4n^2+n$. Thus, the information about $C$ obtained by Alice is bounded by $O(n^2)$. If one views $\mathscr{C}$ as a sequence of random choices from the set $\{H, S, \mathrm{CNOT}\}$ (a universal gate set for Clifford circuits), and each choice of gate is followed another random choice (indexing the input qubit or qubits of the chosen gate) from $\{0, 1, \ldots, n-1\}$ or $\{0, 1, \ldots, n-1\} \times \{0, 1, \ldots, n-2\}$, then the whole information contained in $\mathscr{C}$ can be measured by the entropy of those choices. By Shannon's information theory[25], this entropy equals $L[\log 3 + \log n + (1/3)\log(n-1)]$, where $L$ is the number of gates in $\mathscr{C}$. Therefore, the leakage ratio of the circuit information is bounded by $\frac{4n^2+n}{L[\log 3+\log n+(1/3)\log(n-1)]}$, which is negligible if $L$ is exponential in $n$. □

**Computational complexity and compactness.** In both protocols, the client needs to encrypt her plain data before the homomorphic evaluation and to decrypt the ciphered result after the evaluation (additionally, to measure qubits in the Pauli-$X$ and $Y$ basis in Protocol 1), which is linear in the size of the input; the server performs the homomorphic evaluation, whose computational complexity is the same as that of the non-homomorphic version (up to a multiplicative constant).

In Protocol 1, the communication complexity between the key center and the server is in proportion to the size of the evaluated circuit, so it is not compact. In Protocol 2, the simplified key-update procedure performed by the client is polynomial in the input size of the evaluated circuit $\mathscr{C}$, which is independent of the length of $\mathscr{C}$. Thus, Protocol 2 is compact and is especially suitable for a long evaluation with a short input.

Protocol 2 satisfies perfect security and compactness simultaneously, which is rarely achieved among related works. After this work is completed, we find a similar work in ref. [26], where the server in scheme GT also generates a key-update polynomial $f(x_0, z_0)$ for the final decryption key (when considering only Clifford circuits). However, the formulas for this polynomial is not presented in detail and the calculation rule for $\{p_j\}_{j=0}^{2n-1}$ in Protocol 2 is derived independently.

In Table 3, we give a comparison between our protocols and other closely related schemes. The abbreviations CI, QI, and q-IND-CPA is short for classical interaction, quantum interaction, and indistinguishability under chosen plaintext attack by quantum polynomial-time adversary. The variable "$R$" is the number of $T$ gates in the circuit to be evaluated. For a detailed definition of "$R^2$-quasi-compact", please refer to Definition 9 in Section 3.4 of ref. [11].

## Conclusion

In this paper, we have presented a homomorphic quantum search protocol on encrypted superposition states based on QHE. To homomorphically perform the Grover's algorithm successfully, the QHE schemes utilizing CHE are ruled out. Instead, a trusted third party is introduced to communicate with the search server and update the key, which greatly alleviates the client's workload. This idea is not only applicable to homomorphic quantum search but also adapted to delegated quantum computing wherein the computing purpose is easy to describe but the computing process is long and complex. Moreover, this search protocol can be combined with other delegated quantum computing schemes, so that general tasks including normal computing as well as quantum search can be accomplished homomorphically. Besides, we have provided a non-interactive QHE protocol for Clifford circuits, which is both secure and compact. Based on a new security definition for QHE schemes with TTP, we have proved that the first protocol satisfies this security. We have also defined two kinds of circuit privacy and proved that they are fulfilled by the two protocols, respectively.

## Data availability

All data generated or analysed during this study are included in this published article (and its Supplementary Information Files).

## Acknowledgements

We gratefully acknowledge this work is supported by the Science and Technology Program of Shenzhen of China under Grant Nos. JCYJ20180306124612893. The 5th author also gratefully acknowledges the support from the China Postdoctoral Science Foundation under Grant No. 2017M620322.

## Author contributions

Q.Z. and S.L. proposed the idea, Y.C. designed the quantum circuits, L.L. and J.S. analyzed the security. Q.Z. wrote the manuscript. All authors discussed and reviewed the manuscript.

## Competing interests
The authors declare no competing interests.

## Additional information
**Supplementary information** is available for this paper at https://doi.org/10.1038/s41598-020-61791-9.

**Correspondence** and requests for materials should be addressed to S.L. or Y.C.